\documentclass[floatfix,aps,pra,twocolumn,showpacs,superscriptaddress,10pt]{revtex4-2}

\usepackage[utf8]{inputenc}
\usepackage[english]{babel}

\usepackage{mathtools}
\usepackage{float}
\usepackage{siunitx}
\usepackage{amsmath}
\usepackage{amsfonts}
\usepackage{amssymb}
\usepackage{graphicx}
\usepackage[caption=false]{subfig}
\usepackage{color}
\usepackage[colorlinks=true,citecolor=blue,linkcolor=blue,urlcolor=blue]{hyperref}
\usepackage{times}
\usepackage{amstext}
\usepackage{latexsym}
\usepackage[running,mathlines]{lineno}
\usepackage{verbatim}
\usepackage{physics}
\usepackage{bm}
\usepackage{ulem}
\usepackage{qcircuit}
\usepackage[version=3]{mhchem}
\usepackage{lipsum}  

\newcommand{\beq}{\begin{equation}}
\newcommand{\eeq}{\end{equation}}
\newcommand{\barr}{\begin{eqnarray}}
\newcommand{\earr}{\end{eqnarray}}

\begin{document}
\raggedbottom

\title{Alternatives to a nonhomogeneous partial differential equation quantum algorithm}

\author{Alexandre C. Ricardo}
\affiliation{Departamento de F\'{i}sica, Universidade Federal de S\~{a}o Carlos, 13565-905 S\~{a}o Carlos, S\~{a}o Paulo, Brazil}

\author{Gabriel P. L. M. Fernandes}
\affiliation{Departamento de F\'{i}sica, Universidade Federal de S\~{a}o Carlos, 13565-905 S\~{a}o Carlos, S\~{a}o Paulo, Brazil}

\author{Eduardo I. Duzzioni}
\affiliation{Departamento de Física, Universidade Federal de Santa Catarina,
88040-900, Florianópolis, SC, Brazil}

\author{Vivaldo L. Campo}
\affiliation{Departamento de F\'{i}sica, Universidade Federal de S\~{a}o Carlos, 13565-905 S\~{a}o Carlos, S\~{a}o Paulo, Brazil}

\author{Celso J. Villas-Boas}
\affiliation{Departamento de F\'{i}sica, Universidade Federal de S\~{a}o Carlos, 13565-905 S\~{a}o Carlos, S\~{a}o Paulo, Brazil}

\begin{abstract} Recently J. M. Arrazola \textit{et al.} [Phys. Rev. A \textbf{100}, 032306 (2019)] proposed a quantum algorithm for solving nonhomogeneous linear partial differential equations of the form $A\psi(\textbf{r})=f(\textbf{r})$. Its nonhomogeneous solution is obtained by inverting the operator $A$ along with the preparation and measurement of special ancillary modes. In this work we suggest modifications in its structure to reduce the costs of preparing the initial ancillary states and improve the precision of the algorithm for a specific set of inputs. These achievements enable easier experimental implementation of the quantum algorithm based on nowadays technology.

\end{abstract}

\maketitle

\section{Introduction}

Partial differential equations (PDE's) frequently arise in problems related to science and engineering as a good way to describe rate of variations of a physical quantity in space and time. As in many other fields, quantum computing has strongly impacted the study of solutions for differential equations \cite{leyton2008quantum,clader2013preconditioned,berry2014high,montanaro2016quantum,berry2017quantum,Arrazola_2019,linden2020quantum,lloyd2020quantum,Childs2020,xin2020quantum,kolden2020quantum,childs2021high,romeiro2021quantum}. Following breakthroughs provided by quantum computing at other branches of mathematics, such as linear algebra \cite{harrow2009quantum,lloyd2014quantum,lloyd2016quantum,biamonte2017quantum,lloyd2020quantum1}, great efforts have been made in order to develop quantum algorithms able to provide advantages over traditional methods of solving specific types of differential equations like ordinary differential equations \cite{berry2014high,berry2017quantum,xin2020quantum,Childs2020,knudsen2020solving,romeiro2021quantum,zanger2021quantum}, PDE's with periodic boundary conditions \cite{clader2013preconditioned,montanaro2016quantum,linden2020quantum,kiani2020quantum,garciamolina2021solving,childs2021high,pollachini2021,goes2021qboost}, nonlinear differential equations \cite{leyton2008quantum, lloyd2020quantum,knudsen2020solving,kolden2020quantum,goes2021qboost,liu2021efficient}, and nonhomogeneous partial differential equations \cite{Arrazola_2019}. 

Mathematically, one can associate a differential equation with the action of a classical differential operator $A$ on a solution function $\psi(\vb{x})$, transforming it into a nonhomogeneous function $f(\vb{x})$ as $A\psi(\vb{x})=f(\vb{x})$, where $\vb{x}=(x_1, x_2, ..., x_N)$ and $f(\vb{x})$ is a function over $\mathbb{R}^N$. Generally speaking, if $A$ contains derivatives with respect to at least two independent variables of $\psi$ and $f\neq 0$, the differential equation is said to be a nonhomogeneous PDE. Assuming $A$ to be linear, the solution can be split in a homogeneous solution $\psi_H(\vb{x})$ that satisfies $A\psi_H(\vb{x})=0$, defined only by the differential operator $A$ and the boundary conditions, and a particular solution $\psi_p(\vb{x})$ that satisfies $A\psi_p(\vb{x})=f(\vb{x})$. As many problems of practical interest can be formulated as nonhomogeneous PDE's, such as sound, heat, and electric field distributions, there is a huge interest in developing quantum algorithms for solving this specific type of differential equation.

In \cite{Arrazola_2019}, Arrazola \textit{et al.} proposed a quantum algorithm in the context of continuous-variable model of quantum computing \cite{Lloyd_1999} for solving nonhomogeneous PDE's. Similarly to quantum algorithms for solving systems of linear equations \cite{harrow2009quantum}, the Arrazola's algorithm inverts the differential operator $A$ and computes one particular solution $\psi_p(\vb{x}) = A^{-1} f(\vb{x})$, thus providing a state vector $\ket{\psi_p(\vb{x})}$ that is proportional to the solution of the nonhomogeneous equation. The algorithm is suitable for solving PDE's associated to square-integrable nonhomogeneous functions and differential operators that can be represented as Hermitian matrices. The accuracy of the algorithm relies on the precision of the momentum detection and fidelity of the initial state of ancillary modes required in the protocol, since one of the ideal resource initial state is non-normalizable and must be approximated by quantum physical states. 

In this work, we present possible modifications on the structure of Arrazola's algorithm aiming to reduce the costs of preparing the initial ancillary states. For some classes of linear operators $A$ we show that we can even improve the precision of the algorithm for a specific set of inputs. This work is organized as follow: In Section \ref{Algorithm}, the Arrazola's algorithm is reviewed and some difficulties related to the preparation of its initial ancillary states in a possible physical implementation are highlighted. In Section \ref{Proposals}, we present possible modifications in the structure of Arrazola's algorithm aiming at an easier implementation for PDE's associated to nonhomogeneous functions with a specific spectrum. In Section \ref{Examples}, we give examples of applications of the algorithm, comparing the modifications made and its effects over the original version. Our concluding remarks follow in Section \ref{Conclusions}.

\section{Arrazola's algorithm \cite{Arrazola_2019} for solving a nonhomogeneous partial differential equation}\label{Algorithm}

The objective of Arrazola's algorithm is to find a particular solution of a nonhomogeneous PDE by inverting the operator ${A}$ such that ${A}^{-1} f(\vb{x})=\psi_p(\vb{x})$. In order to do this, the algorithm assumes that (i) the differential operator is polynomial in the variables and its derivatives 
and (ii) the nonhomogeneous function $f:\mathbb{R}^N \longrightarrow \mathbb{C}$ is smooth and square-integrable. In what follows, as to propose modifications, we will briefly review the original Arrazola's algorithm, dividing it in two stages to better comprehension: (a) Encoding the inputs and preparing the auxiliary modes and (b) performing Hamiltonian simulation and post-measuring the output. 

\begin{figure}
    \centering
    \includegraphics[scale = 0.25]{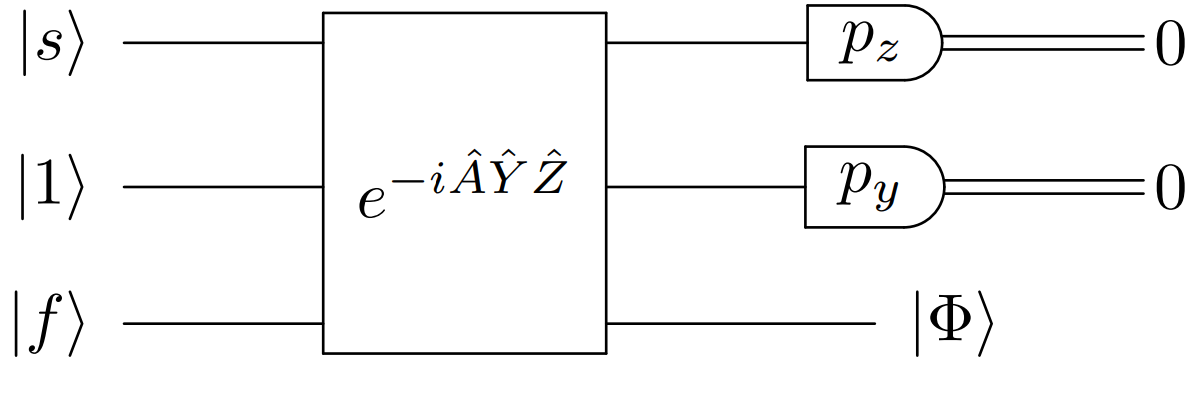}
    \caption{Reproduction of the schematic representation of the quantum algorithm proposed in \cite{Arrazola_2019}. The operators $\hat{A}$, $\hat{Y}$, and $\hat{Z}$ act on the nonhomogeneous state $\ket{f}$, single Fock state $\ket{1}$, and step function state $\ket{s}$, respectively. The system follows homodyne momentum measurements on auxiliary modes ($p_z$ and $p_y$), which after postselecting them on $p_{z,y}=0$ returns the solution of the algorithm $\ket{\Phi} = A^{-1}\ket{f}$.  }
   \label{qcirc1}
\end{figure}

In the first stage, the nonhomogeneous function is encoded in a quantum state $\ket{f}$ of $N$ registers, here considered as being the modes of the quantized electromagnetic field such that $\braket{\vb{x}}{f}\propto f(\vb{x})$. The differential operator $A$ can then be written as a Hamiltonian $\hat{A}$ of the $N$-mode system in terms of the $2N$ available position $\hat{X}$ and momentum $\hat{P}$ mode operators. Besides that, it is necessary to prepare the two-mode initial auxiliary state
\beq
i\sqrt{2\pi}\ket{\tilde{1}}\ket{s} = i\sqrt{2\pi}\ket{\tilde{1}}\int_{-\infty}^\infty \Theta(z)\ket{z}dz,
\eeq
where 
\beq
\ket{\tilde{1}} = \int_{-\infty}^\infty dy y \,e^{-y^2/2}\,\ket{y}
\eeq
is the non-normalized single-photon state and $\ket{s}$ is the ideal step function state, defined by the Heaviside step function $\Theta(z)$. The auxiliary states are prepared in the modes identified by having $\hat{Y}$ and $\hat{Z}$ positions operators, as shown in Fig. \ref{qcirc1}, and the constant $i\sqrt{2\pi}$ has been chosen for convenience in what follows.

In the second stage, an unitary transformation $e^{-i\hat{A}\hat{Y}\hat{Z}}$ is applied on the initial state $\ket{\tilde{1}}\ket{s}\ket{f}$. This transformation is equivalent to evolve the initial state under the Hamiltonian $\hat{H}=\hat{A}\hat{Y}\hat{Z}$ for a unit time. This unitary evolution can be exactly decomposed in gates of a universal gate set~\cite{Kalajdzievski_2019} for some classes of $A$ or approximation methods can be used~\cite{Hatano_2005, Sefi_2011}. After that, the auxiliary modes are projected on the zero momentum states, resulting in the output
\begin{equation}
\ket{\Phi}=\frac{i}{\sqrt{2\pi}}\int_{\mathbb{R}^2} y \,e^{-y^2/2}e^{-i\hat{A}yz} \, \Theta(z) \ket{f}dy\, dz,\label{output}
\end{equation}
whose wave function is the desired particular solution $A^{-1} f$. 
To prove that, decompose $\ket{f}$ in terms of the eigenstates of $\hat{A}$,
\begin{equation}
  \ket{f}=\sum_{a} f(a)\ket{a},
\end{equation}
\noindent where the eigenstates have been indexed by their eigenvalues $a$. To simplify the notation, we are considering a discrete set of non-degenerate eigenvalues, but the generalization to continuous spectrum and/or degeneracy is straightforward. Using this representation in Eq.~(\ref{output}), we get
\begin{equation}
   \label{IdealPhi1}
    \ket{\Phi}=\sum_a \frac{f(a)}{a}\ket{a} = A^{-1}\ket{f}.
\end{equation}
At this stage, the algorithm is not physically possible to be implemented since the Heaviside step function is not square-integrable and the homodyne measurement can not select $p=0$ states perfectly. Thus, Arrazola {\it et al} have investigated modifications to make the algorithm physically possible at the cost of obtaining a less precise particular solution. The step function was replaced by a barrier of length $L$: $b(z)=1/\sqrt{L}$, $0\leq z\leq L$; $b(z)=0$, otherwise, which was then approximated by a truncated expansion of $n$-photon states up to a cutoff dimension $d$. Thus, 
\begin{equation}
   \label{sapprox}
   \ket{s}\to \ket{b_{d,L}} = \frac{1}{\Gamma}\sum_{n=0}^d \gamma_n\ket{n},
\end{equation}
with
\barr
\gamma_n &=& \braket{n}{b_{d,L}} = \frac{1}{\sqrt{L}}\int_0^L \varphi_n(z) dz,\\
\Gamma &=& \sqrt{\sum_n \gamma_n^2},
\earr
where $\varphi_n(z) = C_n H_n(z)e^{-z^2/2}$ is the normalized $n$-photon eigenstate, with $C_n=\big(2^n n! \sqrt{\pi} \big)^{-1/2}$, $H_n(z)$ is the Hermite polynomial of order $n$, and $\Gamma$ is the normalization constant.

The projection of the auxiliary modes on $p=0$ states was replaced by a projection on states of form
\begin{equation}
  \ket{\Delta_{p_y}} =\sqrt{\frac{\Delta}{\pi^{1/2}}}\int_{-\infty}^\infty e^{-y^2\Delta^2/2}\ket{y}dy,
\end{equation}
which have a width $\Delta$ in momemtum space. As $\Delta$ approaches $0$, the momentum detection becomes more selective around $p_{y(z)}=0$. Note that postselecting the outcome of a $\Delta=1$ projection is equivalent to a projective measurement on the vacuum state.

Considering these approximations, the system prepared in the state
\beq
\ket{\Psi} = \lambda \ket{\tilde{1}} \, \ket{b_{d,L}}\,\ket{f},\label{input2}
\eeq
evolves under the Hamiltonian $H = \hat{A}\hat{Y}\hat{Z}$ during a unit of time to the state
\beq
\ket{\tilde{\Psi}} = \lambda \sum_{n=0}^d \frac{\gamma_n}{\Gamma}\int_{\mathbb{R}^2} y e^{-y^2/2} \varphi_n(z) \ket{y}\ket{z} e^{-i\hat{A}yz}\,\ket{f}.\label{input2til}
\eeq
In the following, the auxiliary modes are projected on $\ket{\Delta_{p_y}}\ket{\Delta_{p_z}}$. The prefactor  $\lambda$ is chosen to be 
\beq
\lambda = i\sqrt{\frac{L}{2}}\frac{\Gamma}{\Delta}, \label{lambda}
\eeq
in order cancel out several normalization constants so that, in the limit $L,d\to\infty$ and $\Delta \to 0$,  we recover the same final result as the original algorithm.

\begin{figure}[th]
    \centering
    \includegraphics[scale = 1.1]{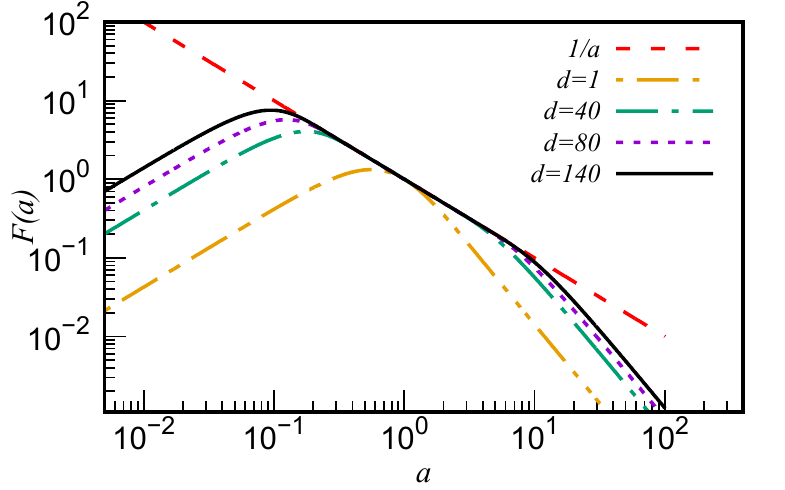}
    \includegraphics[scale=1.1]{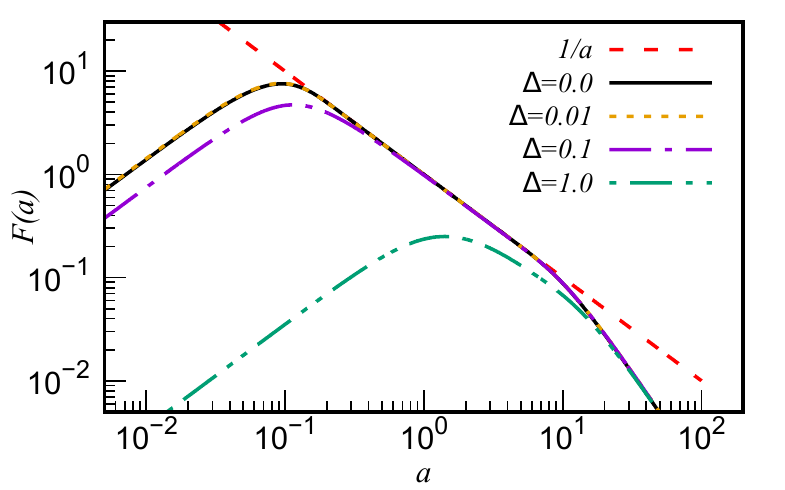}
    \caption{Dilog graphs of $F(a)$ as a function of $a$. Top panel: for different values of $d$, with $\Delta=0$ and $L=20$. Bottom panel: for different values of $\Delta$, with fixed $d=140$ and $L=20$. Both results were obtained through Eq. (\ref{asymptotic1}), which ones were deduced from Ref. \cite{Arrazola_2019}.}
    \label{graph1}
\end{figure}
For realistic values of $L$, $d$ and $\Delta$, the final state is a modified version of that in Eq. (\ref{IdealPhi1}),
\begin{equation}
    \label{approxPhi}
    \tilde{\ket{\Phi}}= A^{-1}_{\rm approx}\ket{f} = \sum_a F(a) f(a) \ket{a},
\end{equation}
where $F(a)\to 1/a$ when $L,d\to \infty$ and $\Delta \to 0$, which corresponds to the ideal solution. To compute $F(a)$, one can expand the state $\ket{f}$ in the eigenstate basis $\ket{a}$ in Eq.~(\ref{input2til}). Without truncating the expansion in Eq.~(\ref{sapprox}), $d \to \infty$, one can evaluate $F(a)$ analytically, whose result is
\begin{equation}
    \label{Fa-dinf}
    F_\infty (a)=\frac{1-e^{-\Big(\frac{a^2}{1+\Delta^2}+\Delta^2\Big)\frac{L^2}{2}}}{\sqrt{1+\Delta^2}}\frac{a}{a^2+\Delta^2(1+\Delta^2)}.
\end{equation}
The above expression is helpful to better understand the effect of the finite length $L$ of the barrier and the finite width $\Delta$. For large enough values of $|a|$, we get $F_\infty(a) \sim 1/a$ as already discussed by Arrazola {\it et al.}~\cite{Arrazola_2019}.

However, when we consider the unavoidable truncation, an important difficulty arises: the correct asymptotic behavior $F_d(a) \sim 1/a$ for $|a|>>1$ is lost. To understand this fact, first note that only odd values of $n$ in Eq.~(\ref{input2til}) will contribute to the final state in (\ref{approxPhi}) (after projecting the auxiliary modes in Eq.~(\ref{input2til}) on $\ket{\Delta_{p_y}}\ket{\Delta_{p_z}}$, due to the odd parity of the state $\ket{1}$, the integration over $y$ leads to an odd function in the variable $z$ that will multiply $\varphi_n(z)$). Neglecting the even Fock states, explicitly writing the odd Fock states' wave functions and integrating over $z$, we obtain
\begin{equation}\label{asymptotic1}
  F(a)=Q(a)\sum_{m=0}^{\left \lfloor{\frac{d}{2}}\right \rfloor} \frac{\gamma_{2m+1}}{\Gamma}h_m(a),
\end{equation}
\noindent where
\begin{equation}
    h_m(a)= \sqrt{\frac{(2m+1)!}{2^{2m+1}}}\frac{(-1)^m}{m!}\Big(\frac{a^2-1+\Delta^4}{a^2+(1+\Delta^2)^2}\Big)^m,
\end{equation}

\begin{equation}
    \label{qa}
    Q(a)=\lambda\,\frac{(-4i\pi^{1/4}\Delta) a}{\big(a^2+(1+\Delta^2)^2\big)^{3/2}},
\end{equation}
and $\lambda$ is given in Eq.~(\ref{lambda}). 
The function $Q(a)$ in Eq.~(\ref{asymptotic1}) behaves asymptotically as $a/|a|^3$ for $|a|\gg 1$. Since each $h_m(a)$ can be written as a power series in $1/a^2$ for large $|a|$, $F(a)$ will behave asymptotically as $a/|a|^3$. This happens even for $\Delta = 0$ and is valid for any set of coefficients $\gamma_n$, not only for those of the finite-length barrier expansion. Therefore, the unexpected asymptotic behavior is entirely due to the truncation. Only an infinite series can fix this problem. In the top panel of Fig.~\ref{graph1}, we compare $F(a)$ for different values of dimension $d$ assuming $\Delta = 0$. We see that $F(1) = 1$ for any dimension $d$. It can be seen directly from Eq.~(\ref{asymptotic1}) that for $a=1$ and $\Delta = 0$, only the first odd state contributes. As the dimension $d$ increases, we see that the range of good convergence to the exact behavior $1/a$ slowly extends both to the left and to the right sides of $a=1$. For $|a|L < 1$, $F(a)$ will depart from the exact behavior for any dimension $d$, while for $|a|\gg 1$, the number of states ($d$) required to have a good approximation may become prohibitively large for practical proposes of physical implementation of the algorithm. In the bottom panel of Fig.~\ref{graph1}, we illustrate the effect of increasing the width $\Delta$, for $L=20$ and $d=140$. One can see that the approximated solutions will be significantly sensitive to the value of $\Delta$, becoming inaccurate when a not small enough width is used, and the use of a small value of $\Delta$ impacts the probability of successfully detect the state $\ket{\Delta_{p_y}}\ket{\Delta_{p_z}}$ on the auxiliary states. The probability of success, given by $\left\|\bra{\Delta_{p_y},\Delta_{p_z}} \ket{\Psi}\right\|^2$, where $\ket{\Psi}$ is the normalized state $e^{-i\hat{A}\hat{X}\hat{Y}}\ket{1}\ket{b_{d,L}}\ket{f}$, can be evaluated as

\begin{equation}\label{P_arrazola}
    P_{0}(\Delta)=\frac{8\Delta^2}{\Gamma^2}\sum_{a}\left|\sum_{m=0}^{\left\lfloor\frac{d}{2}\right\rfloor} \frac{af(a)\gamma_{2m+1}h_m(a)}{\left(a^2+(1+\Delta^2)^2\right)^{3/2}}\right|^2.
\end{equation}

Therefore, the necessary realistic considerations of having a finite-length barrier, $\Delta > 0$ and truncation of the series impose limitations on the accuracy of the physically possible algorithm. Such algorithm will be able to solve accurately the nonhomogeneous PDE only for a function $f$ whose spectrum of eigenvalues $a$ is strongly concentrated around $a=1$. In what follows, we propose two alternative modifications of the practical version of Arrazola's algorithm to recover the correct behavior $F(a) \sim 1/a$ for large $a$, aiming an easier implementation for a more specific class of differential operators.
     
\section{Alternative proposals} \label{Proposals}

\subsection{Proposal 1}

In the last section, we showed that $F(a)$ behaves asymptotically as $a/|a|^3$ for large $|a|$. To recover the exact behavior of $1/a$, our first proposal begins by applying the algorithm to the state $\hat{A}\ket{f}$ instead of the state $\ket{f}$. By doing so, the new modified algorithm would return 
\begin{equation}
  \label{asymptotic2}
  F(a)=\lambda\,\frac{(-4i\pi^{1/4}\Delta) a^2}{\big(a^2+(1+\Delta^2)^2\big)^{3/2}}\sum_{m=0}^{\left \lfloor{\frac{d}{2}}\right \rfloor} \frac{\gamma_{2m+1}}{\Gamma}h_m(a).
\end{equation}
Now $F(a)$ behaves asymptotically as $|a|^{-1}$ for large $|a|$, being insensitive to the sign of the eigenvalue $a$. Accordingly, with the present modification, the algorithm should be used for positive (or negative) semi-definite operators.

Besides changing the initial state from $\ket{f}$ to $\hat{A}\ket{f}$, in our first proposal we also change how the constant $\lambda$ is defined. Thus, in order to recover the $1/\vert a \vert$ behavior for $F(a)$ as discussed above, now we must choose the constant $\lambda$ so that
\begin{equation}
  \lim_{a\to\infty} |a|F(a) = 1.\label{limite}
\end{equation}

In addition, we notice that the general form of $F(a)$ in Eq.~(\ref{asymptotic2}) does not depend on the  specific values of the coefficients $\gamma_n$. Originally, such coefficients were given by the decomposition of the barrier function $b(z)$ in the Fock state basis, but in the present proposal we 
choose those coefficients in a different way. For $|a|\gg 1$, taking into account the series expansion of both the prefactor and $h_m(a)$ in powers of $1/|a|$, we find that $F(a)$ will have the form
\begin{equation*}
F(a) = \frac{1}{|a|}\left(1 + \frac{\beta_1}{a^2} + \frac{\beta_2}{a^4} + \cdots\right).
\end{equation*}
From this, we propose to replace $b_{d,L}(z)$ by an odd normalized function (since only odd $n = 2m+1$ contribute to $F(a)$) such that
\beq
b_{d,L}(z) \to \phi_M(z) = \sum_{m=0}^M \tilde{\gamma}_{2m+1} \varphi_{2m+1}(z),
\eeq
where the coefficients $\tilde{\gamma}_1$,\dots,$\tilde{\gamma}_{2M+1}$ are chosen to systematically result in $\beta_1 = \cdots = \beta_{M} = 0$, so that the error $F(a) - 1/|a| = O\left(1/|a|^{2M+3}\right)$, for large $|a|$.

\begin{figure}
    \centering
    \includegraphics[scale = 1.1]{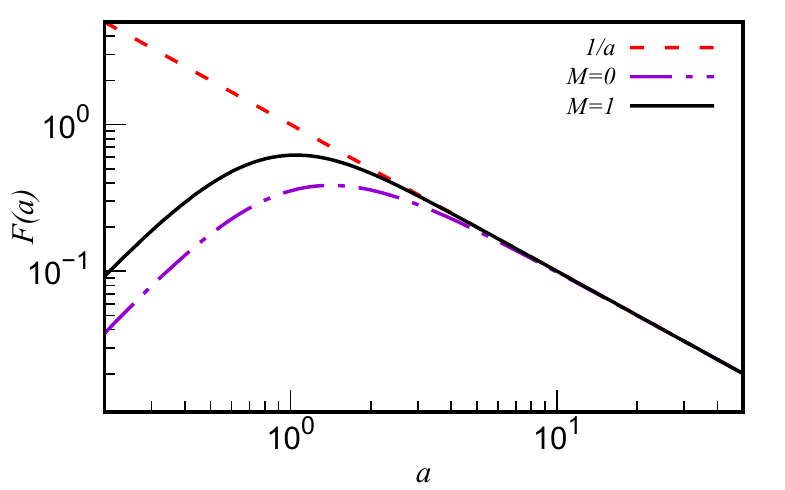}
    \includegraphics[scale = 1.1]{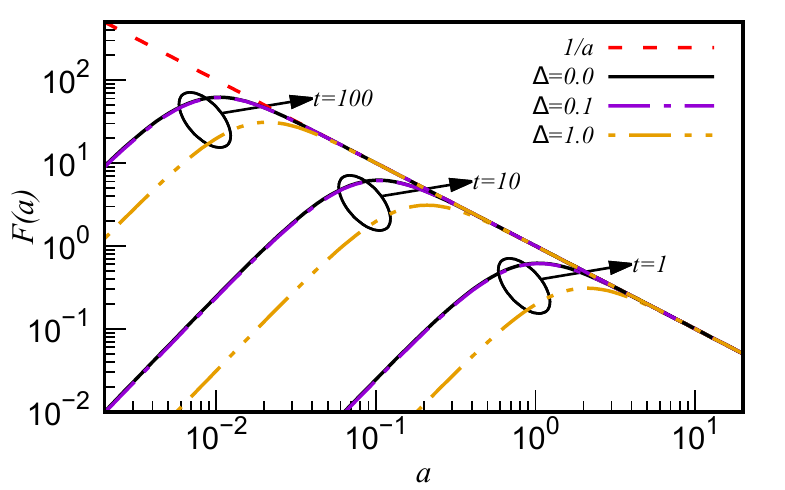}
    \caption{Dilog graph of $F(a)$ as function of $a$ for the first modified algorithm (Top) with $M=0,1$ and fixed $\Delta=0.01$, and (Bottom) for three distinct values of evolution time $t=\{1.0, 10.0, 100.0 \}$ with $\Delta=\{0.01,1.0 \}$. In this case, the projective measurement on the Fock states with postselection on the vacuum state replaces the homodyne detection of the linear momentum with postselection on $p=0$.}
    \label{graph3}
\end{figure}

Fig.~\ref{graph3} shows $F(a)$ obtained according to the present proposal for $M=0$, where $\phi_0(z) = \varphi_1(z)$, and for $M=1$, where $\phi_1(z) = \tilde{\gamma}_1 \varphi_1(z) + \tilde{\gamma}_3 \varphi_3(z)$ with
\beq \label{superp1}
\tilde{\gamma}_1 = \frac{\tilde{\gamma}_3}{\sqrt{6}}\left(3 + \frac{4}{1+\Delta^2 }\right).
\eeq
With two states, the convergence to $1/|a|$ for $|a|>1$ is improved, as expected.
Therefore, the convergence problem for large eigenvalues has been overcome. On the other hand, for $|a|\ll 1$ the results are worse than those using the original version of Arrazola's algorithm. Now, $F(a)\sim a^2$ instead of $F(a)\sim a$ as seen in Fig.~\ref{graph1}. 

The difficulty with small eigenvalues $a$ leads to the last step of our proposal 1. We consider to evolve the initial state for a time $t>1$, through the operator $\exp\left(-i\hat{A}\hat{Y}\hat{Z}t\right)$, instead of keeping $t=1$. Accordingly, $a \to at$ in Eq.~(\ref{asymptotic2}), for example. To keep with Eq.~(\ref{limite}), we have $\lambda\to \lambda t$ also. The effect of using a time of evolution $t>1$ is to assure a good convergence of $F(a)$ to $1/|a|$ for $|a| > 1/t$. Using $t=10$, for example, one gives a result that is similar to the best case seen in Fig.~(\ref{graph1}) on the small $a$ region, without having any problem on the large $a$ region (apart from the restriction to positive(negative)-definite operators). With these considerations, our proposal 1 assumes the final form
\beq
F(a) = \lambda t\,\frac{(-4i\pi^{1/4}\Delta) a^2t^2}{\big(a^2t^2+(1+\Delta^2)^2\big)^{3/2}}\sum_{m=0}^M \tilde{\gamma}_{2m+1}h_m(at).\label{prop1}
\eeq
The time of evolution is a parameter that is naturally at our disposal. Using it only requires that all the gate transformations used in the time-evolution be done in times proportion ally larger. In practice, 
this requires physical systems with longer coherence times.\\

The use of a time of evolution $t>1$ opens the possibility of simplifying the implementation of the algorithm by adopting projective measurements on $\ket{\Delta_{p_y}}\ket{\Delta_{p_z}}$ with width $\Delta = 1$, which corresponds to project on the vacuum states in the auxiliary modes. This is illustrated in Fig.~\ref{graph3}, where the use of $t=10$ and an expansion with two states ($M=1$) allows good convergence with $\Delta = 1$ for $a>0.3$. For practical reasons, projecting on the vacuum-state is convenient due to the easier implementation and because even if states with $\Delta \ll 1$ were implemented, the value of $\Delta$ would likely be subject to experimental uncertainty.\\

Given the normalization constant $\Omega=\|\hat{A}\ket{f}\|$, the probability of successfully detecting the auxiliary modes in the desired states is given by
\begin{equation}\label{P_1}
    P_{1}(\Delta)=\frac{8\Delta^2}{\Omega^2}\sum_a\left|\sum_{m=0}^M \frac{ta^2 f(a) \tilde{\gamma}_{2m+1}h_m(at)}{\left((ta)^2+(1+\Delta^2)^2\right)^{\frac{3}{2}}}\right|^2.
\end{equation}

\noindent This way, the behavior of the probability function is of order $O(\Delta^2)$ for small $\Delta$, similarly to the practical version of Arrazola's algorithm. It can be seen directly from Eq.~(\ref{P_1}) that for fixed evolution times, the probability of success of measuring $\Delta=1$ is higher than for small values of $\Delta$.\\

Before discussing a second alternative of modification of the practical version of Arrazola's algorithm, let us summarize the modifications in our Proposal 1:\\
\indent 1. $\ket{f}\to \hat{A}\ket{f}$;\\
\indent 2. $\lambda$ chosen so that $F(a)$ satisfies Eq.~(\ref{limite});\\
\indent 3. $b_{d,L}(z) \to$ odd function $\phi_M(z)$ to speed up convergence for large $|a|$ and to simplify implementation (no need for large superposition of Fock states);\\
\indent 4. $t > 1$ to improve convergence of $F(a)$ to $1/|a|$ for small $|a|$;\\
\indent 5. Use of $\Delta = 1$ for practical reasons, if $t$ is large enough.

\subsection{Proposal 2}

The asymptotic behavior $F(a) \sim a/|a|^3$ in Eq.~(\ref{asymptotic1}) is due to the truncation of the series expansion of the initial state $\ket{b_{d,L}}$ of one of the ancilla modes and to the odd parity of the initial state $\ket{1}$ of the other ancilla mode. This odd parity is in the root of Arrazola's algorithm~\cite{Arrazola_2019}, which comes from the mathematical identity
\beq
\frac{1}{a} = \int_0^\infty G(ay)dy,
\eeq
valid for an odd function $G(y)$ satisfying $\int_0^\infty G(y)dy = 1$.

However, if we look for a function $F(a)$ that approximates $1/|a|$ instead of approximating $1/a$, one can find an alternative that allow us to keep the initial state $\ket{f}$, modifying only the ancilla states, in contrast to Proposal 1. Given a continuous, square-integrable, even function $H(y)$, with $\int_{0}^\infty H(y)\,dy=1$, we have
\begin{equation}
    \frac{1}{|a|}=\int_{0}^\infty H(ay)dy. 
\end{equation}

Replacing the initial state $\ket{1}$ in the original algorithm by the vacuum state $\ket{0}$, whose wave function attends the conditions of the function $H$ except for a multiplicative constant, we can follow, step by step, the original algorithm. We start from the state
\beq
\ket{\Psi} = \lambda \ket{\tilde{0}}\ket{b_{d,L}}\ket{f},
\eeq
which evolves under the Hamiltonian $\hat{A}\hat{Y}\hat{Z}$ for a time $t$ to 
the state
\beq
\ket{\tilde{\Psi}} = \lambda \sum_{n=0}^d \frac{\gamma_n}{\Gamma}\int_{\mathbb{R}^2} e^{-y^2/2} \varphi_n(z) \ket{y}\ket{z} e^{-i\hat{A}yzt}\,\ket{f}.\label{input3til}
\eeq
After projecting the auxiliary modes on $\ket{\Delta_{p_y}}\ket{\Delta_{p_z}}$, we
get
\beq
\ket{\tilde{\Phi}} = A^{-1}_{\rm approx}\ket{f} = \sum_a F(a) f(a) \ket{a},
\eeq
where
\beq\label{output2}
F(a) = \lambda t\, \frac{2\Delta}{\pi}\frac{1}{\sqrt{a^2t^2 + (1+\Delta^2)^2}}\sum_{m=0}^{\lfloor d/2\rfloor} \frac{\gamma_{2m}}{\Gamma} g_m(at),
\eeq
and
\beq
g_m(a) = \sqrt{\frac{(2m)!}{2^{2m}}} \frac{(-1)^m}{m!}\left(\frac{a^2 -1 +\Delta^4}{a^2 + (1+\Delta^2)^2}\right)^m.
\eeq
Note that only even states contribute to $F(a)$. As expected, we have $F(a)\sim 1/|a|$ for large $|a|$. On the other hand, for $|a|\ll 1$, $F(a)$ approaches a positive constant instead of approaching zero as it happens in Proposal 1 and in the practical version of Arrazola's algorithm.

As in Proposal 1, the constant $\lambda$ will be chosen so that Eq.~(\ref{limite}) is satisfied and the state $\ket{b_{d,L}}$ will be replaced by
$$\ket{\phi_M} = \sum_{m=0}^M \tilde{\gamma}_{2m}\ket{2m},$$
whose coefficients $\tilde{\gamma}_{2m}$ will be chosen to remove the powers $1/|a|^3$,$1/|a|^5$,\dots,$1/|a|^{2M+1}$ of the expansion of $F(a)$ in powers of $1/|a|$. Accordingly, the final form of our proposal 2 becomes
\beq
F(a) = \lambda t\, \frac{2\Delta}{\pi}\frac{1}{\sqrt{a^2t^2 + (1+\Delta^2)^2}}\sum_{m=0}^{M} \tilde{\gamma}_{2m} g_m(at). \label{prop2}
\eeq

\begin{figure}
    \centering
    \includegraphics[scale = 1.1]{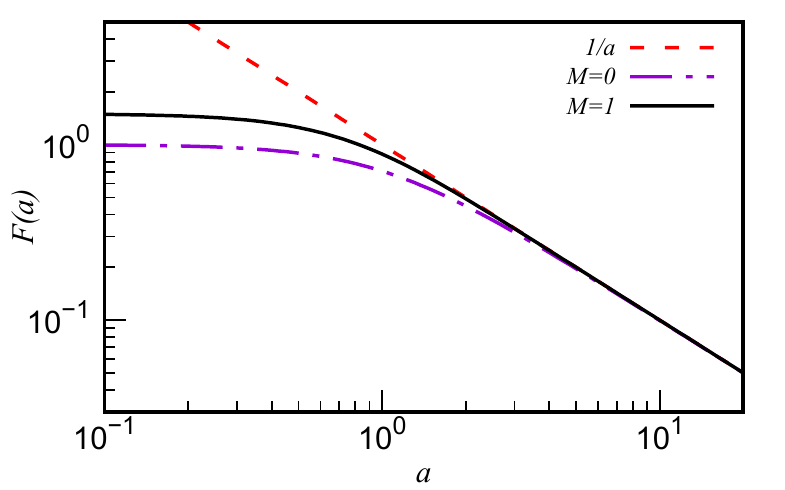}
    \includegraphics[scale = 1.1]{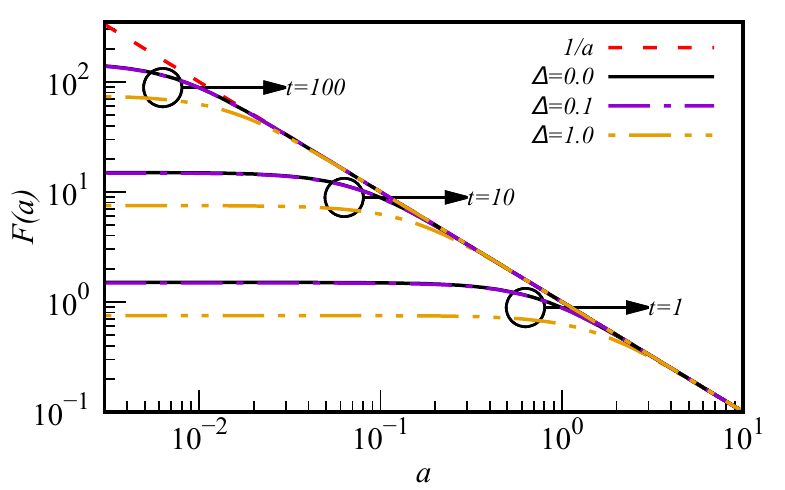}
    \caption{Dilog graph of $F(a)$ as a function of $a$ for the second modified algorithm (Top) with $M=0,1$ and fixed $\Delta=0.01$, and (Bottom) for three distinct values of evolution time $t=\{1.0,10.0,100.0\}$ with $\Delta=\{0.0,0.1,1.0\}$.}
    \label{graph5}
\end{figure}

In Fig.~\ref{graph5}, the top panel shows $F(a)$ obtained according to the present proposal for $M=0$, where $\phi_0(z) = \varphi_0(z)$, and for $M=1$, where $\phi_1(z) = \tilde{\gamma}_0 \varphi_0(z) + \tilde{\gamma}_2\varphi_2(z)$ with
\beq\label{superp2}
\tilde{\gamma}_0 = \frac{\tilde{\gamma}_2}{\sqrt{2}}\left(1 + \frac{4}{1+\Delta^2}\right),
\eeq
evolution time $t=1$ and width $\Delta=0.01$. The results are similar to those of Proposal 1 for $|a|>1$, having a better behavior for $|a|\to 0$ though. In addition, now we have $\ket{f}$ instead of $\hat{A}\ket{f}$ in the initial state. The bottom panel of Fig.~\ref{graph5} illustrates the effect of larger evolution times. The success probability in detecting the states $\ket{\Delta_{p_y}}\ket{\Delta_{p_z}}$ now is given by
\begin{equation}\label{P_2}
        P_2(\Delta)=4\Delta^2\sum_a \frac{|f(a)|^2 \left|\sum_{m=0}^M \tilde{\gamma}_{2m}g_m(at)\right|^2}{(at)^2+(1+\Delta^2)^2}.
\end{equation}

\noindent The behavior of the probability distribution is similar to the cases of Proposal 1 and the practical algorithm, as expected. The modifications of Proposal 2 relative to the practical version of Arrazola's algorithm can be summarized as\\
\indent 1. $\ket{1} \to \ket{0}$ in the initial state;\\
\indent 2. $\lambda$ chosen so that $F(a)$ satisfies Eq.~(\ref{limite});\\
\indent 3. $b_{d,L}(z) \to$ even function $\phi_M(z)$ to speed up convergence for large $|a|$ and to simplify implementation;\\
\indent 4. $t > 1$ to improve convergence of $F(a)$ to $1/|a|$ for small $|a|$;\\
\indent 5. Use of $\Delta = 1$ for practical reasons, if $t$ is large enough.

With large enough evolution times, even the simplest implementation of Proposal 1(2) with the single(zero)-photon state replacing the barrier function and projection on the vacuum states ($\Delta = 1)$ could be used with acceptable accuracy.

\begin{figure}
    \centering
    \includegraphics[width=9cm,scale = 1.1]{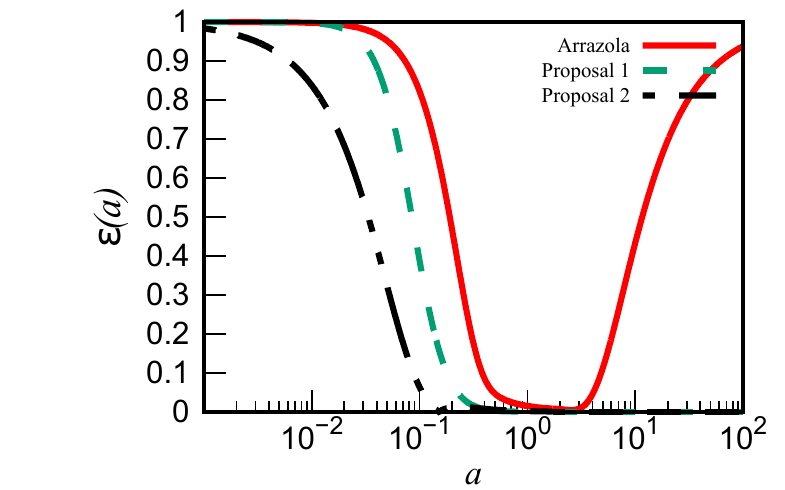}
    \caption{Semilog graph comparing the absolute relative error taking the ideal $1/a$ output as reference, derived numerically using the results of Eqs. (\ref{asymptotic1}) Arrazola's proposal, (\ref{prop1}) Proposal 1 and (\ref{prop2}) Proposal 2 for each proposal with $M=1,\Delta=0.1$ and $t=10$, and the original algorithm for $L=7$, $\Delta=0.1$ and $d=20$.}
    \label{graph6}
\end{figure}

 Fig.~\ref{graph6} compares the relative error, $\epsilon(a)=|1-aF(a)|$, considering similar resources scenarios. In face of this definition, the error behavior of Proposal 2 is slightly better than for Proposal 1 in the region $a>0.1$. On the other hand, with an initial state prepared with a superposition of only $d=20$ Fock states, the original proposal covers a small region with high precision, which is covered by both proposals.

\section{Examples}\label{Examples}

In this section, we present comparisons of the different algorithms discussed
above to solve a nonhomogeneous PDE, that is, the original version of Arrazola's
algorithm and our Proposals 1 and 2. We are considering two examples where the
differential equation is given by a positive definite operator $A$, the Laplace operator in the
first example and a modified version of the quantum harmonic oscillator
Hamiltonian in the second one. Along both examples, we will always adopt $\Delta = 1$ and $M=0$ for Proposals 1 and 2. With this choice of parameters, we have the simplest implementations of those algorithms. The idea is to illustrate that, even with this choice, proposals 1 and 2 can be accurate solvers using  appropriate times of evolution.

\subsection{Laplace operator}

Let us compare the different proposals considering the problem of computing the electrostatic potential $\psi(\mathbf{r})$ associated to a given charge density $\rho(\mathbf{r})$, what requires to solve the Poisson's equation,
\beq
-\nabla^2 \psi(\mathbf{r}) = \rho(\mathbf{r}).\label{poisson}
\eeq
We will consider that the charge density is given by a superposition of Gaussian functions of form 
\begin{equation}
    \label{gaussian}
    g(\sigma,\mathbf{r})= \frac{\sigma^{D/2}}{\pi^{D/4}} \exp{-\sigma^2 r^2/2},
\end{equation}
where $D$ is the spatial dimension and $\sigma$ determines the width of the Gaussian function. The operator $A = - \nabla^2 $ has the plane waves $\ket{\mathbf{k}}$ as eigenstates with eigenvalues $a_{\mathbf{k}} = k^2$,
\beq
A\ket{\mathbf{k}} = k^2 \ket{\mathbf{k}},
\eeq
such that $\braket{\mathbf{r}}{\mathbf{k}} = \exp(i\mathbf{k\cdot r})/(2\pi)^{D/2}$. The state $\ket{f}$,  with wave-function $\rho(\mathbf{r})$, can be written as
\beq
\ket{f} = \int d^Dk G(\mathbf{k}) \ket{\mathbf{k}},
\eeq
where 
\beq
G(\mathbf{k}) = \int \frac{d^Dk}{(2\pi)^{D/2}}\rho(\mathbf{r})e^{-i\mathbf{k\cdot r}}. 
\eeq
A Gaussian density function, $\rho(\mathbf{r}) = g(\sigma,\mathbf{r})$, has a Gaussian Fourier transform,

\begin{equation}\label{gaussian_state}
G(\mathbf{k}) = \frac{1}{\pi^{D/4}\sigma^{D/2}} e^{-\frac{k^2}{2\sigma^2}} = g(\frac{1}{\sigma},\mathbf{k}),
\end{equation}
of width $\sigma$.

For the exact inverse operator $A^{-1}$, we have  $A^{-1}\ket{\mathbf{k}} = 1/k^2 \ket{\mathbf{k}}$, while for each approximate inverse operator $A^{-1}_{\rm approx}$ discussed above, we have
\beq
A^{-1}_{\rm approx}\ket{\mathbf{k}} = F(k^2)\ket{\mathbf{k}},
\eeq
with $F(k)$ determined by the specific algorithm. Accordingly, the exact solution of Eq.~(\ref{poisson}) will be given by
\beq
\psi(\mathbf{r}) = \bra{\mathbf{r}}A^{-1}\ket{f} = \int d^Dk G(\mathbf{k}) \frac{1}{k^2} \braket{\mathbf{r}}{\mathbf{k}},
\eeq
and the approximate solution will be given by
\beq
\psi(\mathbf{r}) = \bra{\mathbf{r}}A^{-1}_{\rm approx}\ket{f} = \int d^Dk G(\mathbf{k}) F(k^2) \braket{\mathbf{r}}{\mathbf{k}}.
\eeq

With the considerations above, let us write explicit expressions for the exact and for the approximate solutions in the case of a Gaussian density, $\rho(\mathbf{r}) = g(\sigma,\mathbf{r})$, in three dimensions. We have
\barr
\psi_{\rm ex}(\mathbf r) &=& \frac{1}{\sqrt{2\pi^{5/2}\sigma^3}}\int_0^\infty \frac{2\sin(kr)}{kr}e^{-k^2/2\sigma^2}\,dk\nonumber\\
&=& \frac{1}{\sqrt{2\pi^{1/2}\sigma^3}} \frac{{\rm erf}(\sigma r/\sqrt{2})}{r}
\earr
for the exact solution, where ${\rm erf}(x)$ is the error function. The approximate solutions will have the general form
\beq
\psi_{\rm app}(\mathbf{r}) = \frac{1}{\sqrt{2\pi^{5/2}\sigma^3}}\int_0^\infty \frac{2\sin(kr)}{kr}\,e^{-k^2/2\sigma^2}F(k^2)k^2\,dk,
\eeq
where $F(a)$ is given by Eq.~(\ref{asymptotic1}) for the practical version of Arrazola's algorithm and by Eqs.~(\ref{prop1}) and (\ref{prop2}) for our proposals 1 and 2, respectively.

\begin{figure}[htp]

\subfloat[Particular solution of the Poisson's Equation (\ref{poisson}), using the parameter $t=1$ in Proposals 1 and 2.]{%
  \includegraphics[clip,width=\columnwidth]{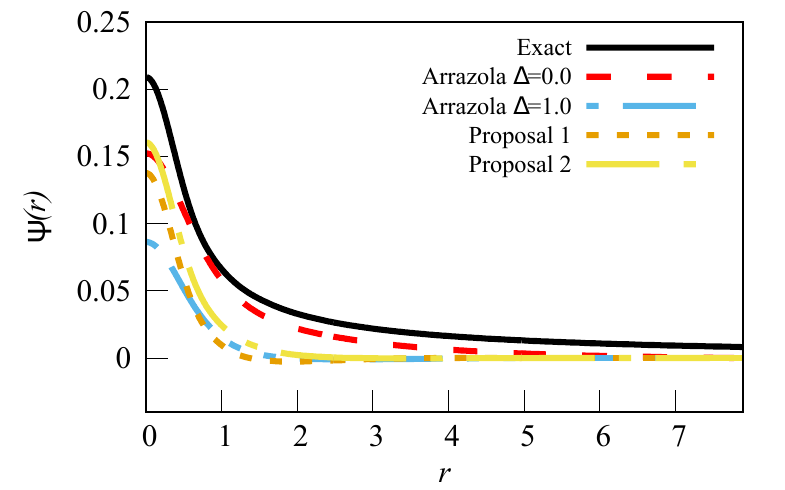}%
}

\subfloat[Particular solution of the Poisson's Equation (\ref{poisson}), using the parameter $t=10$ in Proposals 1 and 2.]{%
  \includegraphics[clip,width=\columnwidth]{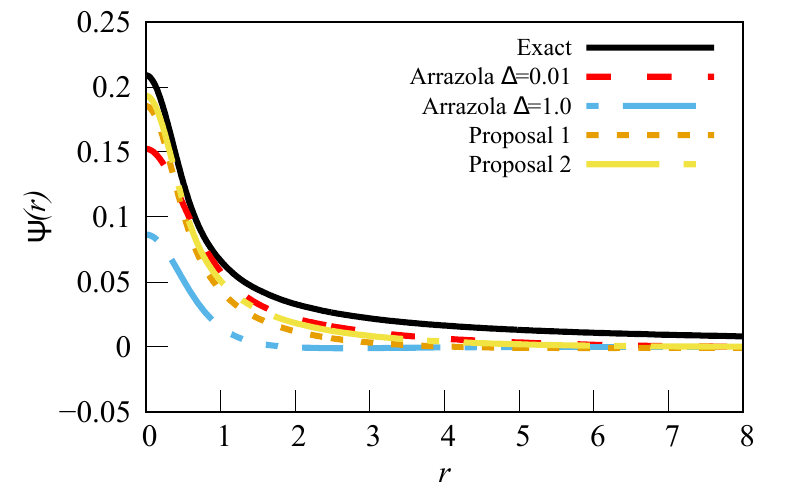}%
}

\subfloat[Particular solution of the Poisson's Equation (\ref{poisson}), using the parameter $t=100$ in Proposals 1 and 2.]{%
  \includegraphics[clip,width=1.0\columnwidth]{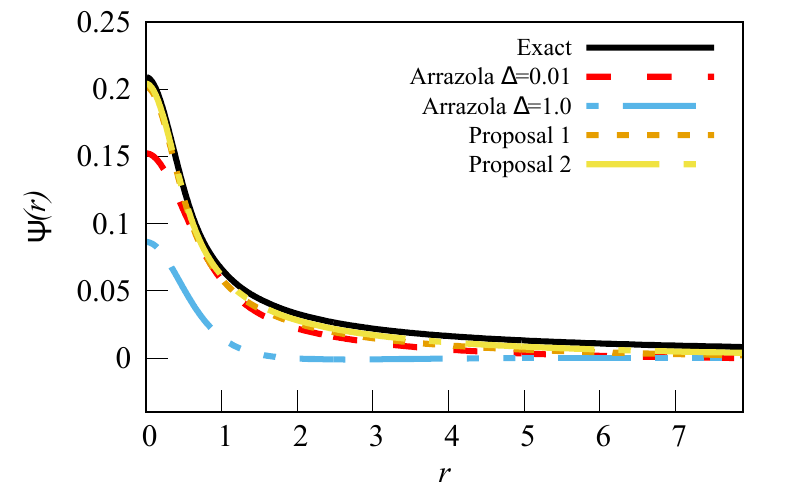}}

\caption{Particular solution of the Poisson's Equation (\ref{poisson}), as a function of the radial variable $r$, with nonhomogeneous function $f(\mathbf{r})=g(4,\mathbf{r})$. The solutions obtained by the original Arrazola's algorithm considering distinct values of $\Delta = \{0.01, 1.0 \}$ are compared to the ones provided by proposals 1 and 2 for the evolution time $t=1.0$ (a), $t=10.0$ (b) and $t=100$ (c)  with $\Delta=1$. The exact output solution is plotted for the sake of comparison.}
\label{sim_sigma4}
\end{figure}

Fig.~\ref{sim_sigma4} compares the solutions obtained by Arrazola's algorithm and by our proposals 1 and 2 with the exact solution of Poisson's equation for a Gaussian charge density with $\sigma=4$. On one hand,  the Fourier transform of such a density has a maximum at $k=0$, so we have a situation where all the approximations will fail in obtaining the correct long range behavior of the exact electric potential, decaying as $1/r$ for large $r$. Obtaining the correct asymptotic behavior would require an exact $F(k^2)$ in the $k\to 0$ limit, exactly where the three approximations are more inaccurate. However, we can see the systematic improvement obtained by increasing the evolution time. For $t=10$, Proposals 1 and 2 are similar to Arrazola's algorithm at large distances. If we had chosen a charge density whose Fourier transform is null at the origin, this problem would not be as severe as it is in this example. On the other hand, $\sigma=4$ implies that the Fourier transform of the charge density is 
appreciable up to $k\sim 4$, what means $a(k) = k^2 \sim 16$. Accordingly, we can see that Arrazola's  algorithm has difficulties to describe the electric potential around the origin, while Proposals 1 and 2 are more accurate for $t=10$ and even more for $t=100$.



\subsection{Quantum Harmonic Oscillator Operator}

Consider the partial differential equation
\begin{equation}
    (-\nabla^2 + \mathbf{x}^2)\psi(\mathbf{x})=f(\mathbf{x}).\label{qhop}
\end{equation}
We will illustrate the different algorithms working with the one-dimensional version of Eq.~(\ref{qhop}).
This ordinary differential equation would arise, for example, when $f(\mathbf{x})$ were a radial function. In the following, we will compare the approximate solutions for
\begin{equation}
    (-\frac{d^2}{dx^2} + x^2)\psi(x)=f(x).\label{qho}
\end{equation}
The differential operator has eigenstates given by 
\begin{equation}
    \label{oscilador}
    \ket{n}=\int \frac{1}{\sqrt{2^n n! \pi^{1/2}}}H_n(x) e^{-x^2/2}\ket{x}dx,
\end{equation}
\noindent with eigenvalues $2n+1$, $n=0,1,2,\dots$, and $H_n(x)$ is the Hermite's polynomial of degree $n$. 
Choosing the coherent state $\ket{\alpha}$ with $\alpha$ real as the state $\ket{f}$, we have \cite{cohen-book} 
\begin{equation}
    \ket{f}=\ket{\alpha}=e^{-\alpha^2/2}\sum_{n=0}^\infty \frac{\alpha^n}{\sqrt{n!}}\ket{n}
\end{equation}
and
\begin{equation}
    f(x)=\frac{1}{\pi^{1/4}}\exp{-(x-\sqrt{2}\alpha)^2/2}.
\end{equation}

We can immediately write the general expression for the approximate solutions to Eq.~(\ref{qho}) that can be obtained by the different versions of the algorithm. We have
\begin{equation}
  \psi(x)=\bra{x}A^{-1}\ket{f(\alpha)}=\frac{e^{-(x^2+\alpha^2)/2}}{\pi^{1/4}}\,S(\alpha,x)
\end{equation}
where
\beq
  S(\alpha,x) = \sum_{n=0}^\infty \frac{1}{n!}\left(\frac{\alpha}{\sqrt{2}}\right)^n\,H_n(x)\,F(2n+1).
\eeq
As in the previous examples, $F(a)$ is given by Eqs.~(\ref{asymptotic1}), (\ref{prop1}) and (\ref{prop2}) for the  original version of Arrazola's algorithm, proposal 1 and proposal 2, respectively. The exact solution is obtained with $F(2n+1) = 1/(2n+1)$.

\begin{figure}[th]
    \centering
    \includegraphics[scale = 1.1]{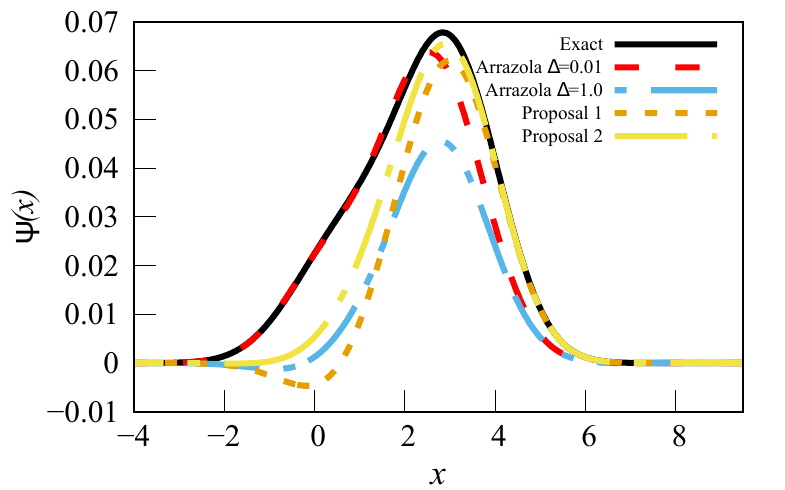}
    \includegraphics[scale=1.1]{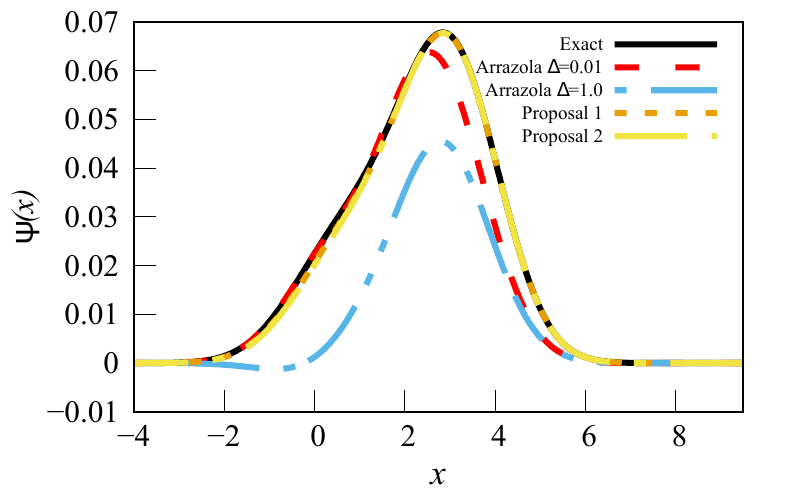}
    \caption{Particular solution of the nonhomogeneous quantum harmonic oscillator equation (\ref{qho}) as function of the position $x$ in which the nonhomogeneous function  $f(x)$ is the wave function of the coherent state with amplitude $\alpha=2.5$. The solutions obtained by the original Arrazola's algorithm considering distinct values of $\Delta = \{0.01, 1.0 \}$ are compared to the ones provided by proposals 1 and 2 for the evolution time $t=1.0$ (top panel) and $t=5.0$ (bottom panel). The exact solution is plotted for the sake of comparison. This result shows that the modified versions of the algorithm have the time evolution as a powerful way to improve its precision.}
    \label{sim_oscillator}
\end{figure}

In Fig. \ref{sim_oscillator}, we compare the original Arrazola's algorithm using $\Delta = \{0.01, 1.0 \}$ with its modifications (proposals 1 and 2) for $\alpha=2.5$, considering two different evolution times, $t = 1.0$ (top panel) and $t=10.0$ (bottom panel). For the short evolution time ($t=1.0$), Arrazolas's algorithm with $\Delta=0.01$ performs better than the alternative proposals for $x < 2$, while for $x>2$ proposals 1 and 2 fit better the exact solution. The problem of Arrazola's algorithm for large eingenvalues $a=2n+1$ of the differential operator, coming from the unavoidable truncation discussed in section \ref{Algorithm}, prevents a satisfactory performance for $x>2$ in this example.  When we increase the time evolution to $t=10$, proposals 1 and 2 become accurate for every value of $x$.

\begin{figure}
    \centering
    \includegraphics[keepaspectratio, width = 9 cm]{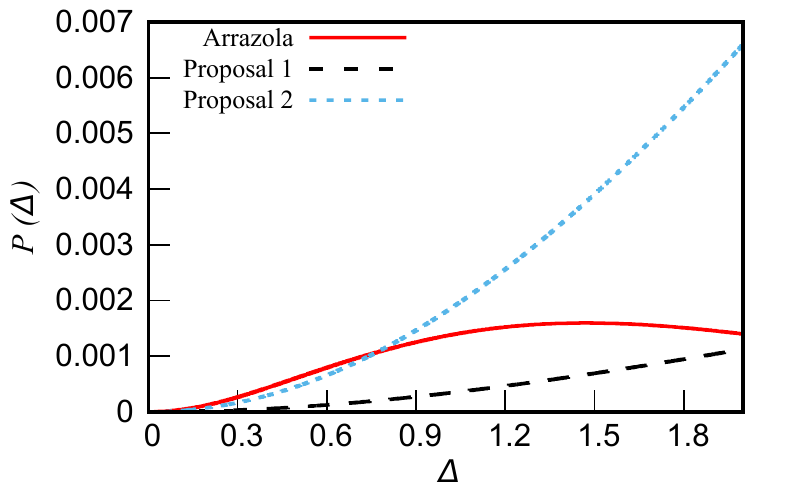}
    \caption{Probability distributions of successfully detecting the state $\ket{\Delta_{p_y}}\ket{\Delta_{p_z}}$ in the auxiliary states for the current example, with $\alpha=2.5$, where the solutions are displayed in Fig.~\ref{sim_oscillator}. The Proposals 1 and 2 are displayed with fixed time $t=10$. In this plot the Proposal 1 is multiplied by a factor $10^3$ and the original algorithm is multiplied by a factor $10$. When $\Delta$ is small, the behavior of the distributions are $O(\Delta^2)$ for the three cases, as expected.}
    \label{prob}
\end{figure}

It is important to notice that the use of $\Delta=1.0$ in the different proposals impact directly the probability of successfully measuring the state $\ket{\Delta_{p_y}}\ket{\Delta_{p_z}}$ on the auxiliary modes. It is direct from Eqs.~(\ref{P_arrazola}), (\ref{P_1}) and (\ref{P_2}) that, when fixing the evolution time, the probability of successfully measuring states with $\Delta=1.0$ is higher than for much smaller values of $\Delta$. In this example, the probability distribution associated to the results of each modification of the algorithm may be seen in Fig.~\ref{prob}.

The original algorithm has a probability of success of \num{3,25e-8} and \num{1,36e-4} for $\Delta=0.01$ and $\Delta=1.0$, respectively, while the probability of success of the first proposal is  \num{3,31e-7} and for the second proposal \num{1,79e-3}, both for $\Delta=1.0$ and $t=5.0$. This shows that the first modification proposal in this example has a probability of success similar to the original algorithm with $\Delta=0.01$ and the second modification proposal has a superior probability of success even for the case where $\Delta=1.0$ in the original algorithm, both having higher precision as displayed in Fig.~\ref{sim_oscillator}.

\section{Conclusions}\label{Conclusions}

In this work, we proposed modifications in the Arrazola's
quantum algorithm for solving nonhomogeneous linear partial differential equations. The modifications aim to improve the precision of the algorithm and to achieve easier experimental implementation by reducing the costs of preparing the required initial ancillary states when dealing with PDE's with semi-definite operators. In this way, our proposals allow to cover a different region, which scales with time evolution, than that where the original algorithm has high precision, leading the way to possible new applications and improvements. We also noted that the error associated to the modified algorithm in Proposals $1$ and $2$ scales slowly 
with the precision of the momentum detection when compared to the original algorithm, which is very useful, especially for platforms where a projective measurement on the vacuum state is more suitable and practical over a homodyne detection, in particular when used along with a longer evolution time so that the error can be mitigated.

\begin{acknowledgments}
This work was supported by the Coordenação de Aperfeiçoamento de Pessoal de  Nível Superior (CAPES)
- Finance Code 001, and through the CAPES/STINT project, grant No.
88881.304807/2018-01. C.J.V.-B. is also grateful for the
support by the São Paulo Research Foundation (FAPESP)
Grant No. 2019/11999-5, and the
National Council for Scientific and Technological Development (CNPq) Grant No. 307077/2018-7. This work is also part of the Brazilian National Institute of Science and Technology for Quantum Information
(INCT-IQ/CNPq) Grant No. 465469/2014-0.
\end{acknowledgments}

\bibliography{refs1}

\end{document}